\documentstyle[twocolumn,prb,aps,epsfig]{revtex} %for preprint version only

\begin{document}
\input{epsf}

\draft
\twocolumn[\hsize\textwidth\columnwidth\hsize\csname@twocolumnfalse\endcsname %for preprint version only

\title{Aging in a topological spin glass}

\author{A.S. Wills$^{1}$, V. Dupuis$^2$, E. Vincent$^2$,
J. Hammann$^2$ and R. Calemczuk$^{1}$ }

\address{$^1$
D\'epartement de Recherche Fondamentale sur la Mati\`ere Condens\'ee,
SPSMS, CEA Grenoble, 38054 Grenoble, France.}

\address{$^2$Service de Physique de L'Etat Condens\'{e}, CEA Saclay, 
91191 Gif sur Yvette, Cedex, France.}

\date{\today}
\maketitle

\begin{abstract}
\leftskip 2.0truecm %for preprint version only
\indent %for preprint version only

We have examined the nonconventional spin glass phase of the
2-dimensional kagom\'e antiferromagnet
(H$_3$O)Fe$_3$(SO$_4$)$_2$(OH)$_6$ by means of {\it ac} and {\it dc}
magnetic measurements. The frequency dependence of the {\it ac}
susceptibility peak is characteristic of a critical slowing down at
$T_g \simeq 18K$. At fixed temperature below $T_g$, aging effects are
found which obey the same scaling law as in spin glasses or
polymers. However, in clear contrast with
conventional spin glasses, aging is remarkably insensitive to
temperature changes. This particular type of dynamics is discussed in
relation with theoretical predictions for highly frustrated  non-disordered 
systems.

\end{abstract}
\vspace{.2in} %for preprint version only


\noindent %for preprint version only
] %for preprint version only
\narrowtext %for preprint version only

\section{Introduction}
\label{sec:intro}

Much recent work has been focused on the nature of glassy magnetic states in  
the absence of quenched disorder.\cite{Bouchaud} 
Simple experimental realizations of non-disordered systems in which
glassy phases have been found are antiferromagnets (AFM) with  kagom\'e and
pyrochlore geometries : (H$_3$O)Fe$_3$(SO$_4$)$_2$(OH)$_6$,
Y$_{2}$Mo$_{2}$O$_{7}$, Tb$_{2}$Mo$_{2}$O$_{7}$,
Y$_{2}$Mn$_{2}$O$_{7}$ and
Yb$_{2}$Ti$_{2}$O$_{7}$.\cite{H3O_Fe,D3O_Fe,D3O_FeAl,Gingras,Gaulin,Hodges}
In the case of the two dimensional (2d) kagom\'e AFM, it is believed
that an {\it xy} anisotropy can induce a finite temperature glass
transition \cite{Spin_fold}. This glassy 
phase has been termed a {\it topological} spin glass, in
order to distinguish it from conventional site-disordered
systems. While the high
degeneracy of these lattices is expected to be robust against the addition
of small amounts of impurities, 
and so should not result in a spin
glass state\cite{Villain}, in some cases conventional behaviour can be
recovered if this dilution is sufficient.
 Examples of this are the 3d pyrochlore Y$_2$Mn$_2$O$_7$ where
the effect of increasing disorder is to progressively restore
conventional spin glass kinetics\cite{Walton}, and the 2d kagom\'e
(D$_3$O)Fe$_{3-x}$Al$_y$(SO$_4$)$_2$(OD)$_6$ where reduction of the Fe
occupancy to $\simeq 90$ \% has been found
to induce long-range magnetic order\cite{D3O_FeAl}.

At present, very little is known about the
glassy phases found in non-disordered frustrated systems. 
In this letter, we study the $S=5/2$ 2d kagom\'e AFM
(H$_3$O)Fe$_3$(SO$_4$)$_2$(OH)$_6$. Our {\it ac} susceptibility measurements
indicate the critical character of the freezing process around
$T_g\simeq18$ K. We find that while aging effects at a fixed temperature below $T_g$
are the same as in site-disordered spin
glasses, there is a very weak sensitivity of these aging phenomena to
temperature variations what is is in sharp contrast with conventional spin
glasses. We propose to understand the transition found in this
2d system, and the slow dynamics observed, in terms of a
theoretical picture of a spin glass state of the ordered kagom\'e AFM,
based on topological excitations termed {\it spin folds}\cite{Spin_fold}.

Our sample was that used in previous studies of the crystal structure,
susceptibility and $\mu SR$ responses\cite{H3O_Fe,muon}. Neutron
diffraction measurements on the deuterated compound have shown that
the glassy phase involves 2d antiferromagnetic correlations that
saturate to a maximum at low temperature, with a spin-spin correlation
length of $\xi = 19 \pm 2$ \AA.\cite{D3O_Fe} Associated with these
correlations is a $T^2$ specific heat, which contrasts with
the linear dependence of conventional spin glasses\cite{Mydosh} and
suggests the presence of 2d 
antiferromagnetic spin waves.

\section{Nature of the freezing mechanism}
\label{sec:II}
We examined the nature of the freezing transition by measuring the
frequency dependence of the {\it ac} susceptibility at 8 frequencies 
ranging from 0.04 to 80 Hz. The out-of-phase component $\chi''$ was
too weak to allow a precise study, but its overall shape compares well
with spin glasses \cite{Mydosh}.  We have found that the temperature of
the $\chi'$-peak shifts very slowly with frequency
($T_{f}(80 \mbox{ Hz})=18.5$ K, $T_{f}(0.04 \mbox{ Hz})=18.0$ K). 
Quantitatively, the
 shift amounts to $\Delta T_f /(T_f\ \Delta
\log \omega)\simeq 0.008$, which is in the same range as
obtained for the critical behaviour of conventional 3d spin glasses
\cite{Tholence,Mydosh}. As a confirmation, a fit of the $T_f(\omega)$
data to an Arrh\'enius function $\omega=\omega_0 \exp(-U/k_BT_f)$
yields an unphysical value of $\omega_0\sim 10^{130}$ Hz (in
agreement with the previous interpretation of results from the
deuterated sample in
terms of a Vogel-Fulcher law \cite{D3O_Fe}).

The critical character of the freezing transition can also be
confirmed by 
a dynamic scaling analysis of the same data. The fit to the critical scaling law
\cite{Mydosh} $\omega=\omega_0 [(T_f(\omega)-T_f(0))/T_f(0)]^{z\nu}$
yields, fixing $z\nu=7$, the plausible values of $T_f(0)=17.8$ K and
$\omega_0=5.9\ 10^{11}$ Hz. This transition temperature is far below
the Curie-Weiss temperature, which is here $\Theta _{CW}=-1200\pm 200$ K,
emphasizing a strong influence of frustration. 
In Ref.\cite{Spin_fold}, it has been proposed that, in the presence of
some {\it xy} anisotropy, the kagom\'e  
AFM experiences a Kosterlitz-Thouless type transition to a glassy state.
The prediction of 
an upper estimate of the transition temperature
$T_{KT}\simeq
\Theta _{CW}/48$, obtained from the spin-wave stiffness in the {\it
xy} limit, agrees well
with the experimental transition temperature, as here 
$\Theta _{CW}/48=25$ K $\sim 17.8$ K.

\begin{figure}[p]
\centerline{\hbox{\epsfig{figure=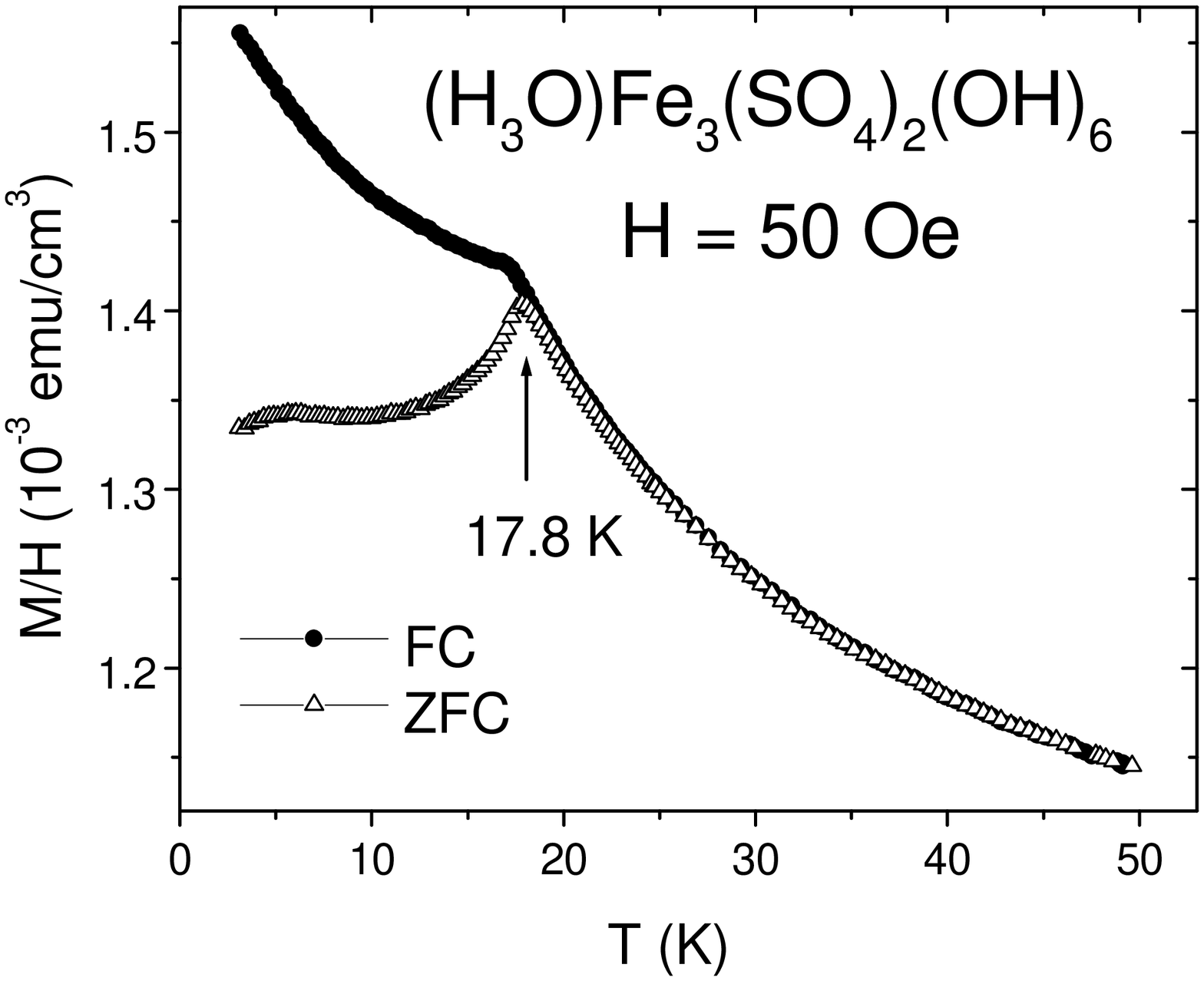,width=7.2cm}}}
\caption{{\it DC} magnetization as a function of temperature.}
\label{Fig1}
\end{figure}

Zero-field cooled (ZFC) and field-cooled (FC) {\it dc} magnetization measurements 
show another aspect of the freezing
transition (Fig.1).  The ZFC-curve peaks at approximately
$T_{ZFC}=17.8$ K ($\simeq 
T_f(\omega \rightarrow 0$), but slight irreversibilities are still
visible as a ZFC-FC separation up to $\sim 25$ K, indicating that some
freezing occurs well above the transition. Also, instead of
the low-temperature FC plateau that is usual in conventional spin
glasses \cite{Mydosh}, there is below $T_{ZFC}$ a rise in the FC
magnetization towards low temperature. Both these
features point towards a progressive freezing of some
super\-paramagnetic entities over a wide temperature
range \cite{superpara}. These species are presumed to be the cause of the slight influence of 
the thermal history between 18 and 25 K on the slow dynamics below
$T_g$ (see Sect. \ref{sec:B}), and their fluctuations are the probable origin of
the dynamic component observed in the $\mu SR$ measurements at low
temperature.\cite{muon} This progressive freezing is superimposed on the critical
behaviour at the transition, and does not prevent the transition from
taking place. 

In a study of other jarosite samples\cite{Earle}, which show stronger super\-para\-magnetic
components, the divergence of the non-linear susceptibility, normally associated with a
spin-glass transition, could not be clearly characterized.

The origin of superparamagnetism at low temperatures in this kagom\'e
system is an open question. Although analytic calculations have shown
that dilution of the magnetic sites creates superparamagnetic
components at high temperatures\cite{Moessner}, it is not clear
whether the presence of such superparamagnetic entities in the spin
glass state is intrinsic to the (2d) kagom\'e lattice, or is the
result of disorder, despite its small level in our sample  (the
occupancy of the magnetic sites, determined from neutron diffraction, is
97.3 (6) \%\cite{H3O_Fe}).

\section{Aging phenomena}
\label{sec:III}
\subsection{Spin glass-like behavior at constant temperature}
\label{sec:A}

In site-disordered spin glasses, the out-of-equilibrium dynamics is
characterized by aging effects: the response to a small magnetic
excitation depends on the time (`age') spent after the quench from
above the freezing temperature \cite{Lundgren,agingAC,gene,old87}.  As
the {\it ac} signal of our present sample is small,
we have explored the nonequilibrium dynamics by means of
{\it dc} magnetization relaxation. In such measurements, aging results
in the dependence of the response to a small field variation, applied
after waiting a time $t_w$ from the quench, on two time variables:
$t$, the observation time (i.e. the time from the field
variation), and $t_a=t_w+t$, the total aging time \cite{gene}. In our
experiments the sample was first cooled from 25 to 12 K in a field of
$50$ Oe. Then, after waiting a time $t_w$ at 12 K, the field was
removed, and the slow decay of the `thermo-remanent' magnetization
(TRM) was recorded as a function of $t$.

\begin{figure}[p]
\centerline{\hbox{\epsfig{figure=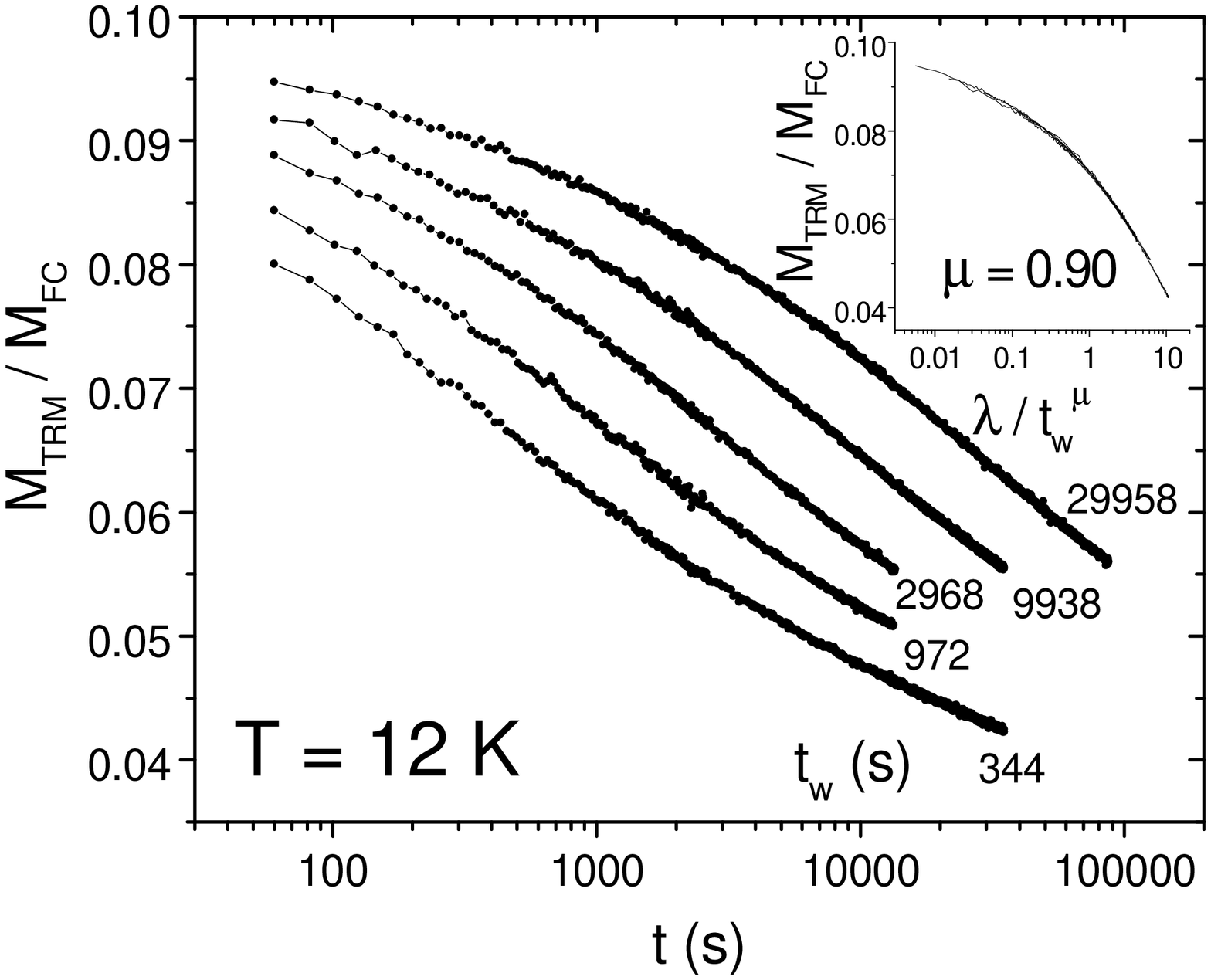,width=8.0cm}}}
\caption{Decay of the thermo-remanent magnetization at $12$ K, 
for various values of the waiting time $t_w$. The
inset shows the scaling of the same data as a function of the
time reduced variable \cite{scalingSG} $\lambda/t_w^\mu$, with $\mu=0.90$ .}
\label{Fig2}
\end{figure}

The observed behaviour, shown in Fig.\ref{Fig2}, is precisely the same
as that of site-disordered spin glasses\cite{gene}: the longer $t_w$, the slower the decay
of the TRM. The system has become `stiffer' with time. The curves
also present the commonly observed inflection point in the region of $\log
t\sim \log t_w$ \cite{Mydosh,Lundgren}. The $t_w$-dependence of these
responses satisfies the same scaling properties as site-disordered
spin glasses\cite{scalingSG} and glassy polymers\cite{Struik}, as shown in the inset of Fig.\ 2.

\subsection{Weak sensitivity to temperature variations}
\label{sec:B}

%\subsection{Aging with positive thermal cycles}
%\label{sec:D}

In site-disordered spin glasses, temperature cycling protocols have been developed in
order to characterize the influence on aging of small temperature
variations (remaining below $T_g$). They have shown, in contrast with the
 expectations for thermally activated processes, that a small 
{\it positive} temperature cycle can erase previous
aging (`rejuvenation' or `chaos-like' effect) instead of simply speeding up
the evolution \cite{old87,agingAC,gene}.

We have investigated the effect of such temperature variations on the
 jarosite system.  The procedure of {\it positive} temperature cycling
 is sketched in the inset of Fig.\ 3. After aging $9700$ s at $T=12$
 K, the sample is heated to $T_0 +\Delta T$ during a short time ($\sim
 100$ s). Following this cycle, aging at $12$ K is continued during
 $300$ s, and the field is cut off (at $t=0$). Fig.\ 3 shows the
 comparison of the resulting decay curves with the reference TRM
 curves obtained after an {\it isothermal} aging of $\sim 300$ s or
 $10000$ s at $12$ K.

\begin{figure}[p]
\centerline{\hbox{\epsfig{figure=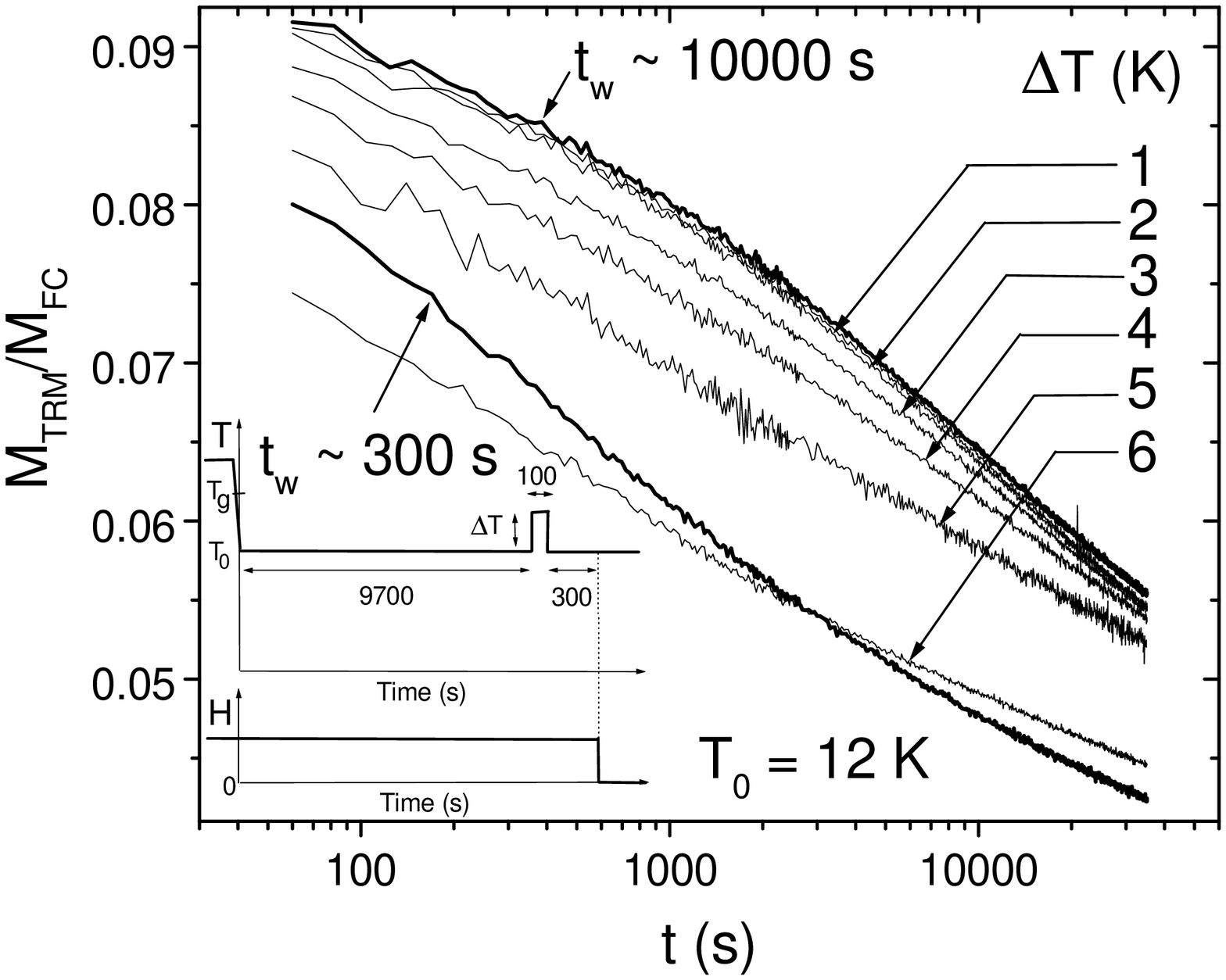,width=8.0cm}}}
\caption{Effect on the TRM-decay at $T_0=12$ K of a {\it positive} temperature
cycling of amplitude $\Delta T$ (thin lines). $T_g=17.8$ K corresponds
to $\Delta T=5.8$ K. 
 Reference curves
(isothermal aging during $t_w\sim 300$ and $10000$ s at $T_0$) are also shown as thick lines.}
\label{Fig3}
\end{figure}

All curves obtained after the positive cycling are {\it lower} than
the $t_w\sim 10000$ s isothermal reference: cycling to $T_0 +\Delta T$ has {\it
not} increased the age of the system. This result also
differs {\it strongly} from the site-disordered spin glass case, in which (in an equivalent
experiment) $\Delta T = 2.5$ K has been found to completely erase previous aging
\cite{old87,gene}.  Here, erasure only occurs when $T_0 +\Delta T$
crosses $T_g$ (for $\Delta T=6$ K), 
and even then results in a curve with a weaker slope than that of the
isothermal $t_w\sim 300$ s
reference. We note that aging at temperatures $T\ge T_g$ still results in some
slowing down of the dynamics, as heating up to $\sim 25$ K is needed to
fully reinitialize aging (at this time scale).

For intermediate $\Delta T$ values ($2-5$ K), Fig.3 shows a partial
erasure of previous aging (but the influence is again much weaker than in conventional spin
glasses). The effect on the curves 
is the same as in site-disordered spin glasses \cite{old87,gene},
partial erasure corresponds to a stronger effect on the short-time
part of the curves (which approaches the `younger' $t_w\sim 300$ s
reference), while the long-time part is less affected (and remains 
close to the `older' $t_w\sim 10000$ s reference).

%We have also checked the effect of a longer duration of the heating
%cycle ($3000$ s instead of $100$ s), which is rather small. For
%$\Delta T=4$ K, the curve remains close to the $t_w=10000$ s
%reference, but its inflection point is now shifted to shorter times
%(`younger' shape: {\it some} aging has been reinitialized).

%\subsection{Aging with negative thermal cycles}
%\label{sec:E}

Negative temperature cycling experiments in
site-disordered spin glasses have shown that the stage of aging before
the temperature cycle can be {\it memorized} during the lower-temperature aging,
in such a way that aging can restart from the same
point when the system returns to temperature $T$.  That is to say, a long aging at $T_0-\Delta
T$ can be of no influence on the effective age at $T_0$, even for {\it
small} $\Delta T$'s ($1-2$ K) for which
little thermal slowing down 
is expected. This {\it memory} effect is equivalent to a growth of
free energy barriers as the temperature is lowered
\cite{agingAC,gene,old87}.

\begin{figure}[p]
\centerline{\hbox{\epsfig{figure=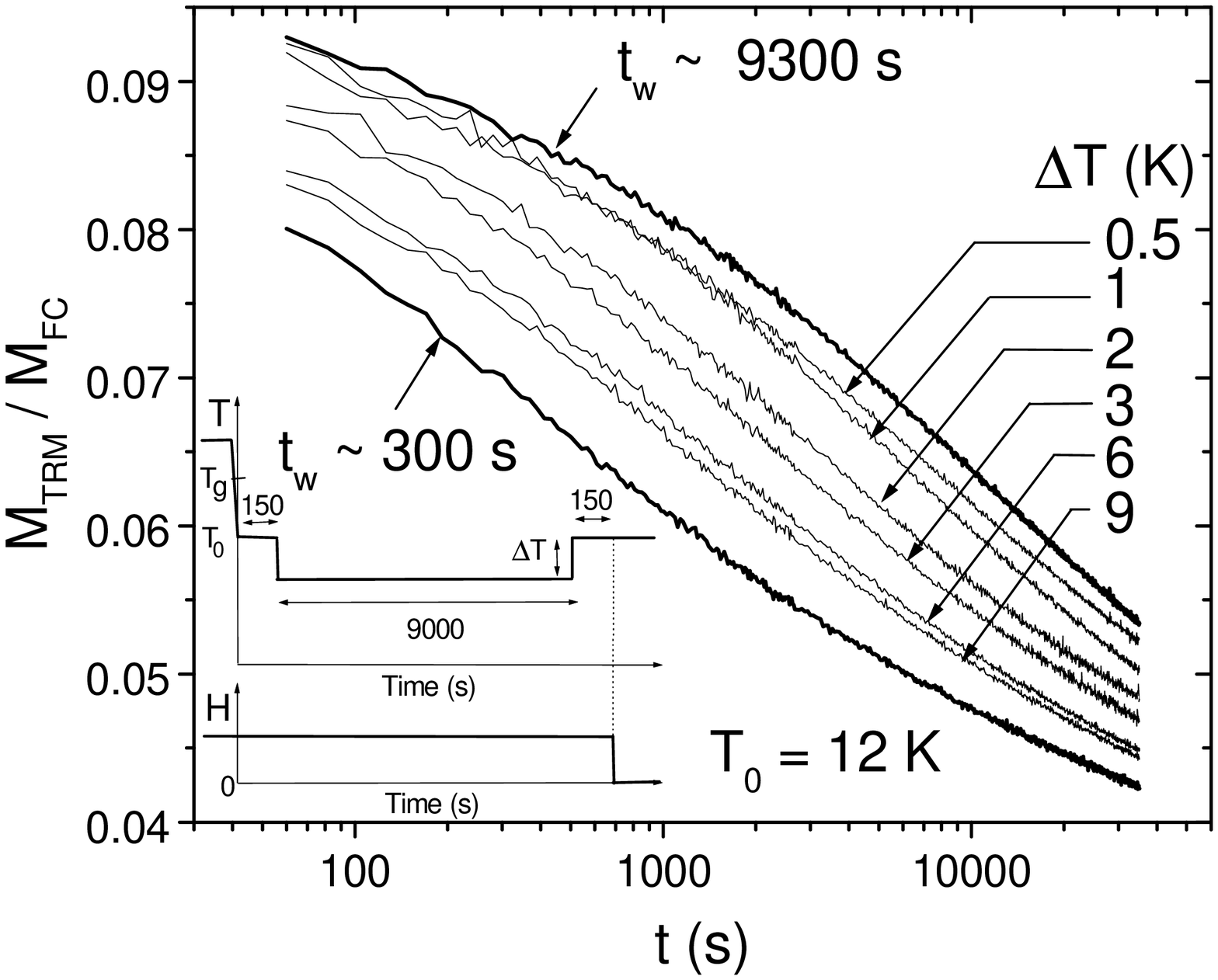,width=8.0cm}}}
\caption{Effect on the TRM-decay at $T_0=12$ K of a negative temperature
cycling of amplitude $\Delta T$ (thin lines). Reference curves
(isothermal aging during $t_w$ at $T_0$) are also shown as thick lines.}
\label{Fig4}
\end{figure}

When applied the negative cycling protocol does not yield these simple memory effects 
and confirms the surprisingly weak influence of
temperature changes on aging,  already observed in the positive cyclings of
Fig.\ 3. This data are shown in Fig.\ 4. 
Since the resulting curves  have
approximately the same shape as those obtained after {\it isothermal}
aging, they can be ascribed an {\it effective
waiting time}, by application of the scaling procedure evoked above
\cite{scalingSG} (this is not the case for the `mixed age' curves
obtained after {\it positive} cyclings). 
Similar (but slightly lower) values of the effective age can be
directly estimated from the position of the inflection point. Qualitatively we find 
that larger values of $\Delta T$
lead to a younger effective age (less influence of aging at $T_0-\Delta
T$). However, {\it quantitatively}, the freezing effect is much weaker
than expected from thermal slowing down, and hence is minuscule when compared 
with the freezing processes of site-disordered spin glasses\cite{agingAC,gene,old87}. 

For $\Delta T=9$ K, aging at
$T_0-\Delta T=3$ K is still seen to add a contribution to aging at
$T=12$ K (the curve is clearly `older' than the $t_w\sim 300$ s
reference). Simple arguments of thermal 
slowing down with fixed energy barriers for such a temperature change, would lead to 
the expectation of even astronomic waiting times having no effect. In
other words, in terms of thermal activation, the observed influence at
$12$ K of aging at $3$ K corresponds to that of a much higher effective
temperature (of the order of $11$ K).

\section{Discussion}
\label{sec:IV}

Thus, we have found that the slow dynamics and aging effects in this
kagom\'e antiferromagnet are, {\it at constant temperature}, similar
to those of site-disordered spin glasses, but the r\^ole played by
temperature changes is markedly different. Positive and negative
cycling experiments have shown that aging is neither reinitialized nor
drastically frozen by temperature changes: it is mostly unaffected, as
long as $T_g$ is not crossed. This behavior is in marked contrast with
conventional site-disordered spin glasses \cite{Jarray}.

%The `non-slowing down' of processes when the temperature is lowered is
%evocative of an evolution through an extremely degenerate ground state
%manifold. 

In a ground state
configuration of the kagom\'e AFM, the nearest-neighbour moments are related by 120
$^{\circ}$, and the local chirality of the triangular motifs is either
uniform or staggered. The domain walls that separate regions of uniform and
staggered chirality correspond to topological defects which cost zero
internal energy. 
In Ref.\cite{Spin_fold}, the transition of the kagom\'e AFM to a
glassy state is described 
in terms of a Kosterlitz-Thouless type mechanism involving the
binding of these defects.  
It has been argued \cite{Spin_fold}
that the processes of defect propagation are `non-Abelian', {\it i.e.}
their formation does not commute. This creates a further hindrance to
the evolution towards a given spin configuration and means that defect propagation 
is not simply a matter of thermally activated barrier
crossing.

While at a given instant the kagom\'e system can be thought of as a
mosaic of chiral domains, the equilibrium state itself is believed to involve
fluctuating short-range chiral correlations\cite{Berlinsky}. The
existence of aging effects is in agreement with Monte Carlo
studies\cite{Berlinsky} and indicates that this state is approached by
slow relaxation and a gradual evolution of the system. The equilibrium state may 
correspond to a temperature dependent distribution of
domains. They will have different Boltzmann weights because those with uniform 
chirality develop extended defects, whereas those within domains of 
staggered chirality are localized \cite{Berlinsky,Spin_fold}. The weak effect of
temperature changes suggests that this distribution of chiral ordering
varies only slowly with temperature.

In conclusion, we believe that
(H$_3$O)Fe$_3$(SO$_4$)$_2$(OH)$_6$  is  
representative of the new class of `topological' spin glasses. 
There is good evidence from {\it ac} measurements that a 
critical transition to a glassy state occurs at $T_g\ne 0$. While its aging dynamics 
at a given temperature follows that of conventional spin 
glass systems, there is a remarkable insensitivity to temperature 
changes that demonstrates clearly the different natures of site-disordered and 
such topological spin glasses.

We acknowledge enlightening discussions with D. Grempel and
 L.F. Cugliandolo.

%\newpage


\begin{references}
\bibitem{scalingSG}In spin glasses (and polymers \cite{Struik}), a universal
relaxation curve is obtained as a function of the 
$\lambda/t_w^\mu$ scaling variable. The time variable
$\lambda=[t_w^\mu/(1-\mu)][(t_w+t)^{1-\mu}-t_w^{1-\mu}]$ accounts for a
logarithmic shift of response times which is proportional to
$(t_w+t)^\mu$, with $\mu$ slightly lower than 1, a general feature of
aging in spin glasses (see details in Ref.\ \cite{gene}).
\end{references}
\end{document}